\documentclass[10pt,twocolumn,superscriptaddress,english,10pt,prl,showpacs,floatfix,aps]{revtex4-2}
\usepackage[latin9]{inputenc}
\usepackage{amsthm}
\usepackage{amsmath}
\usepackage{amssymb}
\usepackage{esint}
\usepackage{graphicx}
\usepackage{upgreek}
\usepackage{siunitx}

\makeatletter
\@ifundefined{textcolor}{}
{%
 \definecolor{BLACK}{gray}{0}
 \definecolor{WHITE}{gray}{1}
 \definecolor{RED}{rgb}{1,0,0}
 \definecolor{GREEN}{rgb}{0,1,0}
 \definecolor{BLUE}{rgb}{0,0,1}
 \definecolor{CYAN}{cmyk}{1,0,0,0}
 \definecolor{MAGENTA}{cmyk}{0,1,0,0}
 \definecolor{YELLOW}{cmyk}{0,0,1,0}
}

\usepackage{soul}
\usepackage{xspace}
\usepackage{braket}
\usepackage[backref=true,
bookmarksnumbered=true,
bookmarks=true,
bookmarksopen=true,
colorlinks=true,
citecolor=blue,
linkcolor=blue,
anchorcolor=green,
urlcolor=blue,unicode=false]{hyperref}
\renewcommand{\v}[1]{\ensuremath{\mathbf{#1}}} % for vectors
 
\let\baraccent=\= % rename builtin command \= to \baraccent
\renewcommand{\=}[1]{\stackrel{#1}{=}} % for putting numbers above =

\newcommand{\mos}{$\text{MoS}_\text{2}$}

\newcommand{\didv}{d$I$/d$V$}

\makeatother

\usepackage{babel}

\begin{document}

\title{%Impact of the moir\'e structure of MoS$_2$/Au(111) on spin excitations of individual Fe atoms\\
Moir\'e tuning of spin excitations: Individual Fe atoms on MoS$_2$/Au(111)}

\author{Sergey Trishin}
\affiliation{\mbox{Fachbereich Physik, Freie Universit\"at Berlin, 14195 Berlin, Germany}}

\author{Christian Lotze}
\affiliation{\mbox{Fachbereich Physik, Freie Universit\"at Berlin, 14195 Berlin, Germany}}

\author{Nils Bogdanoff}
\affiliation{\mbox{Fachbereich Physik, Freie Universit\"at Berlin, 14195 Berlin, Germany}}

\author{Felix von Oppen}
\affiliation{\mbox{Dahlem Center for Complex Quantum Systems and Fachbereich Physik, Freie Universit\"at Berlin, 14195 Berlin, Germany}}

\author{Katharina J. Franke}
\affiliation{\mbox{Fachbereich Physik, Freie Universit\"at Berlin, 14195 Berlin, Germany}}

\begin{abstract}
Magnetic adatoms on surfaces are exchange coupled to their environment, inducing fast relaxation of excited spin states and decoherence. These interactions can be suppressed by inserting decoupling layers between the magnetic adsorbate and the metallic substrate. Using low-temperature scanning tunneling microscopy and spectroscopy, we investigate the magnetic properties of single Fe atoms on Au(111) covered by a monolayer of \mos\ and explore the influence of the moir\'e structure on the spin excitation spectra. Fe adatoms absorbed at minima of the moir\'e structure exhibit almost pure inelastic spin excitations. Kondo correlations develop for adsorption sites off the moir\'e minimum, culminating in fully developed Kondo resonances on moir\'e maxima. The increase in Kondo correlations is accompanied by a decrease in the single-ion anisotropy, which we rationalize by poor-person's scaling. Local variations in the tunneling spectra indicate pronounced interference 
between tunneling paths involving the spin-carrying orbitals and a hybrid Fe--S orbital, which forms despite the nominally inert nature of the terminating S layer.

\end{abstract}

%\pacs{75.75.-c,74.20.-z,75.70.Tj,73.63.Nm}

\maketitle

%\section{Introduction} 
Single magnetic atoms on surfaces hold substantial promise for spin state manipulation and coherent control of quantum spin states \cite{Yang2019}. To protect the spin state against decoherence, exchanges of energy and angular momentum with the substrate need to be suppressed \cite{Willke2018}. While tunneling currents allow for addressing individual spins, 
metallic substrates are problematic as the creation of electron-hole pairs and spin-spin scattering involving the conduction electrons provide highly efficient pathways for energy and angular momentum relaxation \cite{Khajetoorians2011, Khajetoorians2013, Willke2018}. These processes can be efficiently suppressed when using materials with an energy gap around the Fermi energy, either by passivating a metallic substrate with nitride \cite{Hirjibehedin2007} or oxide \cite{Heinrich2004, Loth2010, Paul2017} layers, or by using a superconductor \cite{Heinrich2013}. 

Monolayers of van der Waals materials have recently been proposed as interlayers for decoupling magnetic adsorbates from metallic substrates. Lifetimes of excited spin states as well as coherence times could be extended using hexagonal boron nitride \cite{Kahle2012, Jacobson2015} and graphene \cite{Donati2013, Dubout2015}. Monolayers of transition metal dichalcogenides are another promising class of van der Waals interlayers. 
Each sheet consists of transition metal atoms sandwiched between chalcogen layers with all covalent bonds saturated within the layer. 
Monolayers of \mos\ have been shown to decouple organic molecules from metal substrates \cite{Krane2018, Reecht2019, Reecht2020}. Their decoupling efficiency has been attributed to the electronic bandgap in combination with the layer thickness and the van der Waals nature. These properties should also be favorable for decoupling magnetic atoms from metal substrates. \mos\ layers on Au(111) substrates offer further tunability due to the moir\'e structure resulting from a mismatch between the chalcogen and Au lattice constants. This imposes long-range modulations of the electronic structure and thus of the coupling between adsorbate and substrate \cite{Krane2018a}. Previous work has shown that the moir\'e structure of hexagonal boron nitride (hBN) on Rh(111) induces small variations in the magnetic excitation energies of hydrogenated Co adatoms with adsorption site \cite{Jacobson2015}. A strongly site-specific Kondo resonance of Co atoms can be induced on Ru(0001) covered by graphene, which partially detaches due to periodic rippling \cite{Ren2014}

Here, we exploit a monolayer of \mos\ on Au(111) as a decoupling layer for iron (Fe) atoms. Using scanning tunneling microscopy and spectroscopy at a temperature of \SI{1.1}{\K}, we identify the adsorption site of the Fe adatoms and detect their magnetic fingerprint in tunneling spectra. We show that the data are compatible with a spin state of $S=1$ and sizable axial anisotropy. The moir\'e structure of the \mos\ overlayer on Au(111) imposes strong variations of the exchange coupling between Fe adatoms and substrate. When probing directly above the Fe atoms, we observe almost pure inelastic spin excitations in the minima of the moir\'e structure. Away from the minima, an increase of the moir\'e-modulated density of states enhances the exchange coupling, culminating in a fully developed Kondo resonance for Fe atoms located at moir\'e maxima. Beyond these dramatic variations with adsorption site, there are pronounced local variations in the immediate vicinity of the Fe atom, which we trace to interference of tunneling paths proceeding via different orbitals.

\begin{figure}[tb]
\includegraphics[width=0.5\textwidth]{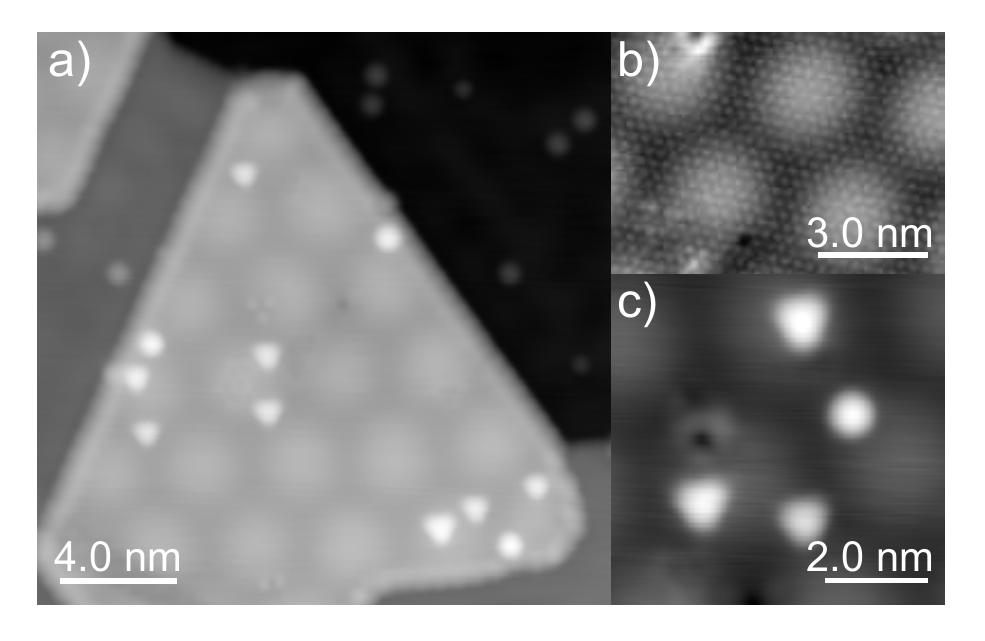}
\caption{(a) Large-scale STM image of a monolayer island of \mos\ on Au(111) after deposition of a small number of iron atoms at low temperature. The atoms are statistically distributed over the bare Au(111) surface and the \mos\ islands. Image recorded at $V=\SI{50}{mV}$, $I=\SI{100}{\pA}$. (b) Atomic-resolution topography of the S lattice on a clean \mos\ island. Atomic defects are seen in the upper and lower left corners of the image. Image recorded at $V=\SI{5}{mV}$, $I=\SI{42}{\nA}$. (c) Closeup of individual Fe atoms on a \mos\ island. At low bias voltages, most atoms appear triangular. Image recorded at $V=\SI{50}{mV}$, $I=\SI{100}{\pA}$.}
\label{mos}
\end{figure}

Figure\,\ref{mos}a shows an STM image of a monolayer island of \mos\ on Au(111) with a small coverage of Fe adatoms. The hexagonal corrugation within the \mos\ island with a periodicity of $\sim$\,3.3\,nm reflects a moir\'e structure reflecting a lattice mismatch between the Au and S layers \cite{Gronborg2015,Bana2018, Krane2018a}. An atomically resolved image of a clean \mos\ area (Fig.\,\ref{mos}b) shows the terminating S layer and resolves a minute amount of intrinsic point defects (see upper and lower edge of the image). We focus on adatoms far from the defect sites to avoid any influence on the magnetic properties.   
 
A close-up view of several Fe adatoms on a \mos\ island is shown in Fig.\,\ref{mos}c. When scanned at low bias voltage, most Fe adatoms appear with a triangular shape, while a few atoms are imaged as a round protrusion (see SM). All triangles point in the same direction relative to the underlying S lattice. The symmetric shape suggests that these Fe atoms are located in hollow sites of the S layer. There are two distinct hollow adsorption sites on the \mos\ islands, with and without a Mo atom underneath. The uniform orientation of the triangular shapes indicates that only one of these is occupied. This inference is corroborated by superimposing a hexagonal lattice representing the S sites on the Fe-covered \mos\ layer. The positions and orientations of the triangular-shaped Fe atoms are all consistent with equivalent hollow sites.

\begin{figure*}[t]
\includegraphics[width=\textwidth]{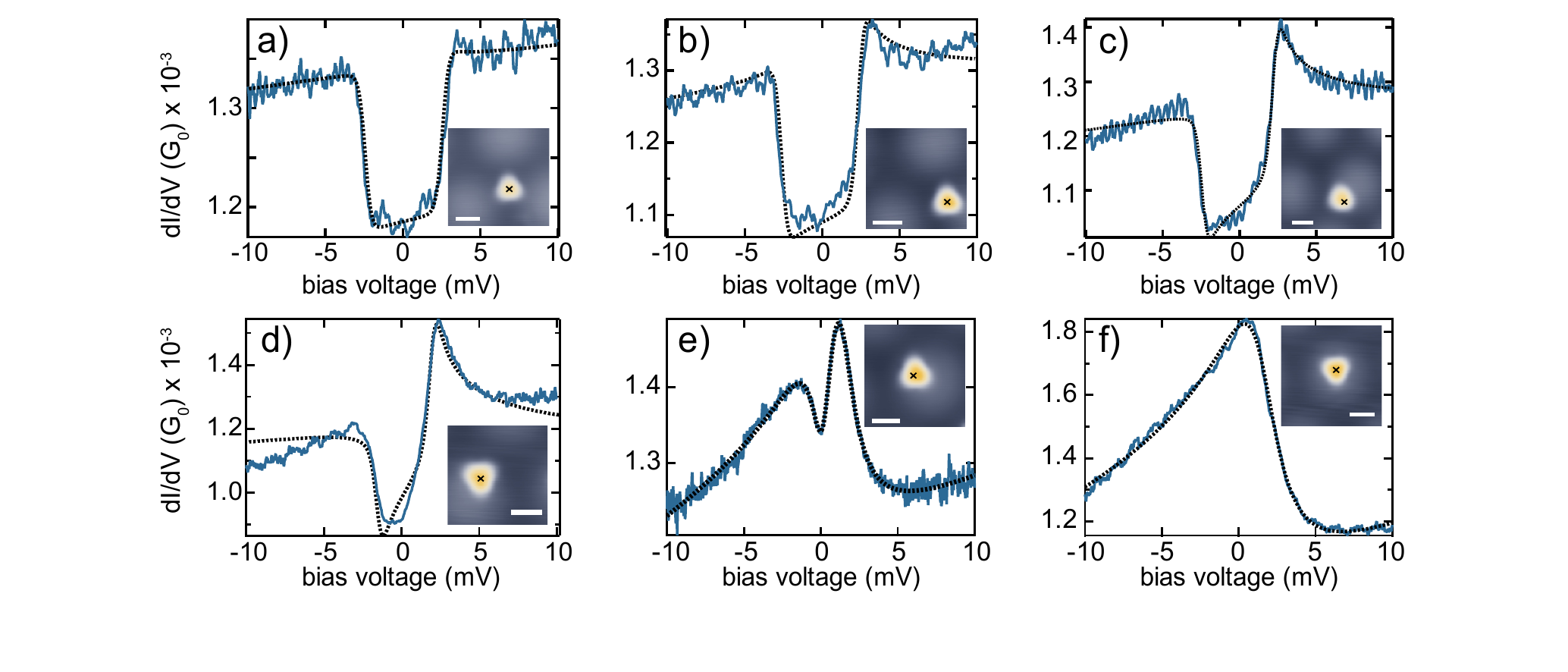}
\caption{Differential conductance spectra taken at the center above individual Fe adatoms (blue lines) in different local environments as indicated by the STM topographies in the insets (scale bars are $\SI{1}{\nm}$, the crosses mark the location of the spectra). Fits are shown as black dashed lines. The spectrum in (a) can be reproduced by symmetric Fermi-Dirac functions with a width of $0.45\pm 0.07 $\,meV. The spectra in (b), (c) and (d) exhibit additional cusps on top of the inelastic steps. These spectra are fitted within a perturbative approach in the exchange interaction $\frac{1}{2}J \widehat{\sigma} \cdot \widehat{S}$, with $J$ the exchange coupling strength and $\widehat{\sigma}$ the vector of Pauli matrices \cite{Ternes2015}. Additional potential scattering $U$ leads to bias-asymmetric lineshapes. The fit uses the code described in \cite{Ternes2015}. Fit parameters: $J\rho=-0.11$, $U=-0.32$, $D=\SI{2.5}{\mV}$ (b), $J\rho=-0.16$, $U=-0.71$, $D=\SI{2.3}{\mV}$ (c) and $J\rho=-0.33$, $U=-1$, $D=\SI{1.73}{\mV}$ (d). Spectrum (e) exhibits a Frota lineshape with a dip at the Fermi level. We fitted the spectrum using a Frota lineshape with an additional Lorentz peak. Fit parameters: Frota phase $\phi = 3.87$, Frota width $\Gamma=\SI{2.21}{mV}$, Frota center $x_0 =\SI{1.3}{mV}$, Lorentz width $\Gamma_L=\SI{0.79}{mV}$ and Lorentz position $x_L =\SI{0.11}{mV} $ In (f), the spin excitation gap has closed, resulting in a fully developed Frota lineshape. Fit parameters: Frota phase $\phi = 4.12$, Frota width $\Gamma=\SI{3.49}{mV}$, Frota center $x_0 =\SI{1.6}{mV}$. The spectra were recorded at a setpoint of $V=\SI{10}{mV}$, $I=\SI{1}{\nA}$.
} 
\label{fig:spectra}
\end{figure*}

To investigate the magnetic properties of the individual triangular Fe atoms, we recorded differential conductance (\didv) spectra of $\sim$\,40 atoms. Fig.\,\ref{fig:spectra} displays a characteristic set of spectra. We find dramatic variations of the spectral lineshapes depending on the Fe adatom's position with respect to the moir\'e pattern. The Fe atom in Fig.\,\ref{fig:spectra}a is located in a minimum of the moir\'e structure. Its \didv\ signal exhibits stepwise increases of intensity at $\pm 2.7\,\mathrm{mV}$. The Fe atom in Fig.\,\ref{fig:spectra}b is shifted slightly off the minimum and shows a small overshoot of the differential conductance at positive bias voltages above the inelastic excitation threshold. This overshoot becomes more pronounced for Fe atom located further from a moir\'e minimum (Fig.\,\ref{fig:spectra}c,d). Concomitantly, the step-like feature shifts to lower energies. Adatoms located close to a moir\'e maximum exhibit lineshapes which resemble a Frota peak with an additional dip at the Fermi level (Fig.\,\ref{fig:spectra}e). Directly at the moir\'e maximum, the inelastic gap disappears and gives way to a fully developed asymmetric Frota peak (Fig.\,\ref{fig:spectra}f).

To understand these spectroscopic characteristics of individual Fe atoms, we first discuss their electronic configuration. In the gas phase, Fe atoms host six electrons in the five-fold degenerate $d$ shell leading to a spin state of $S=2$. 
Consistent with the triangular shape of the electronic structure reflected in the STM images (see further discussion below), the hollow adsorption site imposes a trigonal pyramidal crystal field which lifts the degeneracy of the $d$ levels. Simple crystal field considerations suggest a splitting into two sets of doubly degenerate states, one of mostly $d_\mathrm{xz}$ and  $d_\mathrm{yz}$ character and another of $d_\mathrm{x^2-y^2}$ and  $d_\mathrm{xy}$ character, as well as a high-lying nondegenerate state deriving from the $d_\mathrm{z^2}$ orbital. Depending on the ratio between the crystal field splitting and Hund's energy, filling these levels with six $d$ electrons either leads to a spin state of $S=2$ with a singly-occupied $d_\mathrm{z^2}$ orbital, or to $S=1$ with an empty $d_\mathrm{z^2}$ orbital. 

Due to magneto-crystalline anisotropy, the spin ${\bf\widehat S}$ prefers to align along a distinct direction even at zero magnetic field. This is described by the spin Hamiltonian  $\widehat{H}=D \widehat{S}_{z}^{2}+E(\widehat{S}^{2}_{x}-\widehat{S}_{y}^{2}) $, where $D$ is the axial and $E$ the transverse magnetic anisotropy \cite{Hirjibehedin2007}.
Tunneling via a surface-adsorbed Fe atom may thus not only proceed elastically, when leaving the spin projection $m$ unchanged ($\Delta m=0$), but also inelastically when changing the spin projection ($\Delta m= \pm1$). The opening of inelastic tunneling channels results in a stepwise increase of the junction conductance.

The \didv\ spectrum in Fig.\,\ref{fig:spectra}a can indeed be well reproduced by a broadened Fermi-Dirac distribution function at $\pm\SI{2.7}{\mV}$. 
Thus, the spectra probe magnetic atoms subject to a local crystal field characteristic of the adsorption site. The measured width $w=\SI[separate-uncertainty]{0.45(7)}{\meV}$ of the step is larger than the experimental broadening at $T=\SI{1.1}{\K}$, suggesting a lower bound of $\sim$\,$\SI{1}{\ps}$ for the spin lifetime.
Significantly broader steps are observed for Fe atoms in direct contact with a metal substrate \cite{Balashov2009, Khajetoorians2011, Khajetoorians2013,Hermenau2018}. This confirms that a single layer of \mos\ indeed acts as an efficient decoupling layer for the Fe atoms. 

At the same time, the decoupling efficiency depends sensitively on the position of the Fe atom relative to the moir\'e lattice. As we observe lineshapes ranging from inelastic steps to fully developed Kondo peaks, there must be variations in the magnitude of the coupling to the substrate. At and close to the minimum of the moir\'e structure (Fig.\,\ref{fig:spectra}a,b,c), the inelastic steps can be understood by considering the exchange coupling to the substrate perturbatively. To describe the conductance overshoot, one needs to go beyond lowest-order perturbation theory and include processes in second-order Born approximation \cite{Ternes2015}. The asymmetry in the bias polarity can be reproduced when including potential scattering at the impurity in addition to the exchange coupling. Figure \,\ref{fig:spectra}a,b,c include corresponding fits \cite{Ternes2015}. These indicate that both the exchange coupling $J$ and the potential scattering $U$ increase as the adatom location moves off the moir\'e minimum \cite{Ternes2015, Jacobson2015}. In addition, there is a substantial reduction in the longitudinal anisotropy $D$.

For Fe atoms located further from the moir\'e minima, the pronounced changes in the lineshape can no longer be captured within this perturbative framework. The transition to a fully developed Kondo lineshape requires a strong-coupling approach and can be captured within a poor-person's scaling approach \cite{Anderson1970}. In the present case, one should account for the longitudinal anisotropy as well as the higher impurity spin. Allowing for anisotropic exchange couplings $J_\perp$ and $J_z$ and focusing on the simplest spin state $S=1$ consistent with our experimental data, the scaling equations take the form \cite{Zitko2008}
\begin{eqnarray}
\label{eq:scaling1}
\frac{dJ_z}{d\ell}&=&\frac{1}{1-\Delta}J_\perp^2 \\
\label{eq:scaling2}
\frac{dJ_\perp}{d\ell}&=&\frac{1}{2}(1+\frac{1}{1+\Delta})J_\perp J_z \\
\label{eq:scaling3}
\frac{d\Delta}{d\ell}&=&\Delta- (J_z^2-J_\perp^2)\ln 2.
\end{eqnarray}
Here, the exchange couplings are made dimensionless by means of the density of states $\rho$ of the conduction electrons, $\rho J\to J$, the scaling variable $\ell$ parametrizes the bandwidth $E_c$, $\ell = \mathrm{ln}(E_{c0}/E_c)$, with initial bandwidth $E_{c0}$, and $\Delta =D/E_c$ measures the longitudinal anisotropy in units of the bandwidth. 

As the adsorption site moves towards the moir\'e maximum, the associated change in the density of states increases the bare exchange couplings and hence the Kondo temperature. A fully developed Kondo peak is expected once the Kondo temperature exceeds the longitudinal anisotropy $D$. The scaling equations suggest that this effect is reinforced by the renormalization of $D$. For easy-axis anisotropy ($D>0$ and $J_z>J_\perp$) as well as easy-plane anisotropy ($D<0$ and $J_z<J_\perp$), the anisotropy $D$ is reduced in magnitude during renormalization [see the second term on the right hand side of Eq.\ (\ref{eq:scaling3})]. This further favors the crossover to fully developed Kondo correlations and is consistent with the observation that the threshold voltage of spin excitations decreases as the adsorption site approaches the moir\'e maximum. We note that a renormalization of the anisotropy has also been observed in model calculations using the one-crossing approximation \cite{Jacob2018}. 

\begin{figure*}[ht]
\includegraphics[width=.95\textwidth]{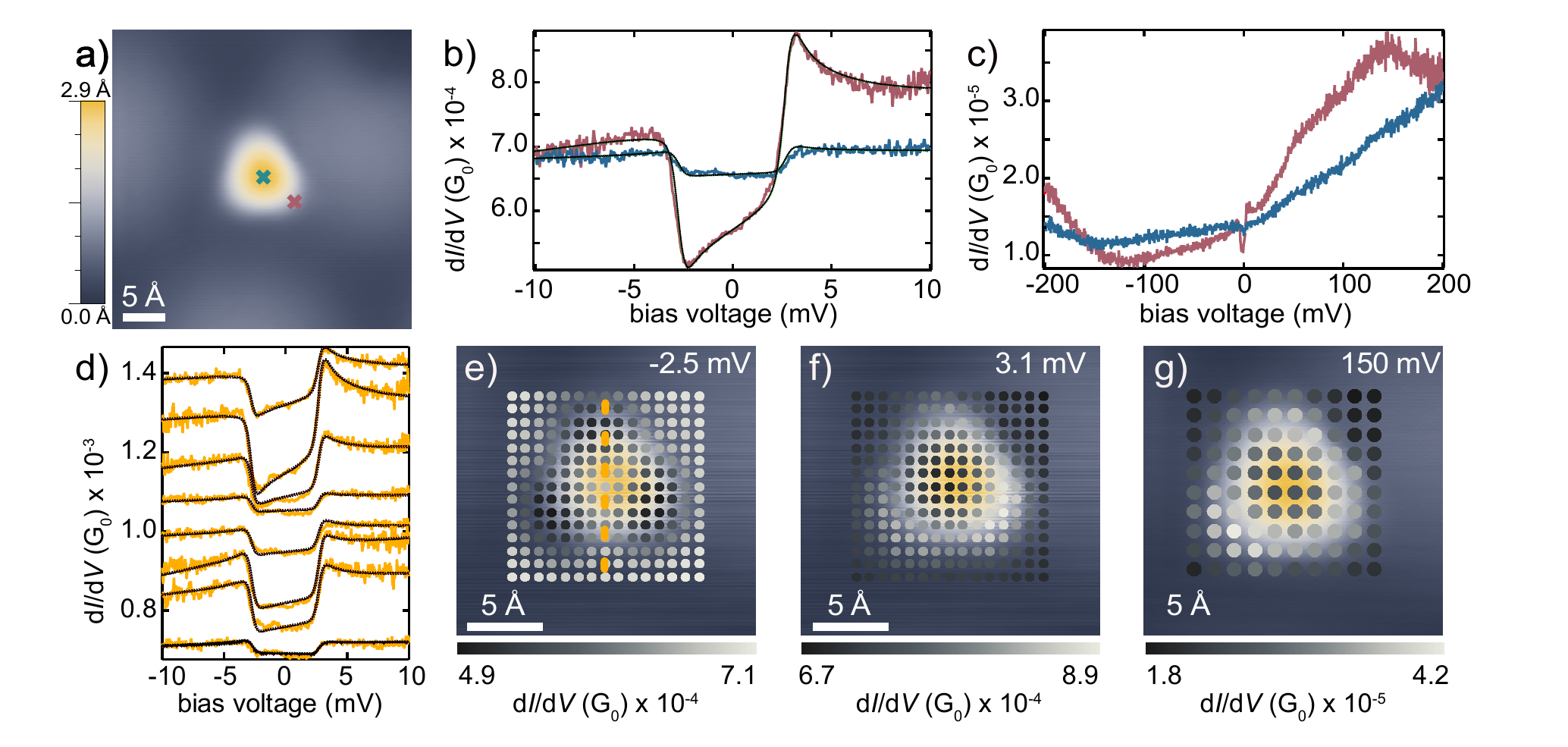}
\caption{(a) STM topography image of an Fe atom located on a moir\'e minimum. (b) Differential conductance spectra at low energies on the center (indicated by blue cross in (a)) and vertex (indicated with a purple cross in (a)) of the triangular shape of the Fe adatoms. Fits as described in \cite{Ternes2015} are shown as black dashed lines.  Fit parameters: $J\rho=-0.13$, $U=-0.14$, $D=\SI{2.75}{\mV}$ (blue) and $J\rho=-0.13$, $U=-0.78$, $D=\SI{2.74}{\mV}$ (red).  (c) Differential conductance spectra over a larger energy range on the center and the vertex of the triangular shape of the Fe adatoms. (d) \didv\ spectra (orange) recorded across the Fe--S complex along the orange line in (e). Fits shown as black dashed lines; $J\rho$ was kept constant in all fits, while $U$ and the transmittivity of the junctions were adjusted. Values of $U$ are shown in SM \cite{SM}. Spectra are offset for clarity. (e-g) STM topographies (blue-yellow, background) with superimposed differential conductance signal at the indicated bias voltage (black-white dots, scale given below panels) extracted from a densely spaced grid of spectra across the Fe atom at the indicated energies. The spectra in (b), (d), (e), and (f) were recorded at a setpoint of $V=\SI{10}{mV}$, $I=\SI{1}{\nA}$ with an additional retraction of the tip by $\SI{20}{pm}$; the spectra in (c) and (g) at $V=\SI{10}{mV}$, $I=\SI{20}{\pA}$ with an additional retraction of the tip by $\SI{20}{pm}$.} 
\label{fig:maps}
\end{figure*}

Beyond the dramatic variations in lineshape as a function of the adsorption site relative to 
the moir\'e lattice, recorded at the centers of the Fe atoms, we observe remarkable spatial variations also in the immediate vicinity of the adatoms. We illustrate these local variations for Fe atoms located at the minimum of the moir\'e structure (Fig.\,\ref{fig:maps}). (Corresponding data for an Fe atom close to the maximum are shown in the SM \cite{SM}.)
At a moir\'e minimum, spectra recorded directly above the center of the Fe atom show pure inelastic steps. Significant additional structure is observed off center near one of the vertices of the triangle (purple cross in Fig.\,\ref{fig:maps}a). Here, the spectrum exhibits substantial bias asymmetry, a conductance overshoot just above the inelastic threshold at positive bias, and an additional dip just below threshold at negative bias (Fig.\,\ref{fig:maps}b). We can fit this spectrum within the perturbative approach \cite{Ternes2015} using the same exchange coupling as for the spectrum taken above the center, but a larger potential scattering amplitude $U$. More generally, all spectra recorded in the immediate vicinity of the Fe atom can be described by adjusting $U$ (as well as the overall tunneling strength) and keeping $J\rho$ constant (see Fig.\,\ref{fig:maps}d for a set of fits for spectra taken along a high-symmetry axis of the Fe--S complex). 

We map out the asymmetry underlying the variations of $U$ by plotting the \didv\ signal at the energies of the dip and the overshoot (Fig.\,\ref{fig:maps}e,f). Both maps reveal that these features are most pronounced at the vertices of the topographic triangular shape (see Fig.\,\ref{fig:maps}a). The variation in $U$ can be understood as an interference effect originating from parallel tunneling paths \cite{Ujsaghy2000,Schiller2000,Frank2015}, similar to tunneling through metal-organic complexes \cite{Verdu2018, Farinacci2020}. To gain insight into the tunneling paths, we record \didv\ spectra for a larger bias-voltage range (Fig.\,\ref{fig:maps}c). Spectra taken at the center of the Fe atoms reveal a wide slope across the Fermi level, but are otherwise featureless, lacking distinct resonances. In contrast, we observe a resonance at  $\sim \SI{150}{mV}$ at the triangle's vertices. This resonance has its largest intensity at the vertices, as seen in the spatial map in Fig.\,\ref{fig:maps}g, suggesting the formation of a hybrid Fe--S state with weight concentrated on the S atoms. The hybridization is sufficiently strong to give the Fe atoms their triangular shape in the STM images, despite the saturated bonds within the \mos\ layer.

The nearly symmetric lineshape at the center of the Fe--S complex indicates a dominant tunneling path with negligible potential scattering, which is due to cotunneling via the spin-carrying orbitals. The more pronounced bias asymmetry at the vertices originates from enhanced potential scattering $U$, which we associate with cotunneling via another orbital. Most likely, this orbital can be identified with the resonance at $\sim \SI{150}{meV}$ as the spatial variations of its map are correlated with the spatial variations of $U$. The total tunneling amplitude emerges from interference between these cotunneling paths. As a function of tip location, their relative contributions vary in accordance with the orbital wave functions, explaining the spatial variations in the potential scattering strength $U$ \cite{Farinacci2020}.

In conclusion, monolayers of \mos\ on Au(111) act as decoupling layers for paramagnetic Fe atoms on metallic substrates. The decoupling efficiency is widely tunable due to the moir\'e modulation of the density of states. As a function of adatom location relative to the moir\'e lattice, we observe lineshapes ranging all the way from pure inelastic excitations to fully developed Kondo resonances. Concomitant with increasing Kondo correlations, the single-ion anisotropy is reduced. Moir\'e structures can thus be employed for tuning the magnetic coupling strength, highlighting the importance of local properties of decoupling layers. Moreover, atomic-scale variations of the tunneling lineshapes originate from inelastic excitations via different orbitals, indicating the formation of Fe--S hybrid states despite the van der Waals character of \mos.

\section{Methods}
We have performed STM experiments on Fe adatoms deposited on single-layer molybdenum disulfide islands grown on a Au(111) single crystal. The experiments were carried out at a base temperature of \SI{1.1}{\K}. The Au single-crystal surface was cleaned by repeated sputter-anneal cycles until a clean, atomically flat surface was obtained. The \mos\ islands were grown by depositing Mo atoms on the Au(111) surface and subsequent annealing to 800 K in H$_2$S gas at a pressure of p=$10^{-5}$\,mbar \cite{Gronborg2015, Krane2018a}. The as-prepared layers were decorated with Fe atoms by evaporation onto the sample held at $\sim \SI{10}{\K}$. Differential conductance spectra were recorded in constant-height mode, using a lock-in amplifier. We used a modulation frequency of $f=\SI{911}{\Hz}$. Modulation voltage and set point conditions are given in the figure captions. The grid of spectra were analyzed using the freeware SpectraFox \cite{SpectraFox}.

\begin{acknowledgments}
We thank David Jacob for discussions and acknowledge financial support by Deutsche Forschungsgemeinschaft through TRR 227, project B05. 
\end{acknowledgments}

\bibliographystyle{apsrev4-2}

%\bibliography{library}

%

\clearpage
\setcounter{figure}{0}
\setcounter{section}{0}
\setcounter{equation}{0}
\setcounter{table}{0}
\renewcommand{\theequation}{S\arabic{equation}}
\renewcommand{\thefigure}{S\arabic{figure}}
	\renewcommand{\thetable}{S\arabic{table}}%
	\setcounter{section}{0}
	\renewcommand{\thesection}{S\arabic{section}}%

\onecolumngrid

\newcommand{\vsigma}{\mbox{\boldmath $\sigma$}}

\section*{\Large{Supplementary Material}}

\maketitle 

\section{Identification of Fe adatoms on defects}

\begin{figure}[htb]
\includegraphics[width=\textwidth]{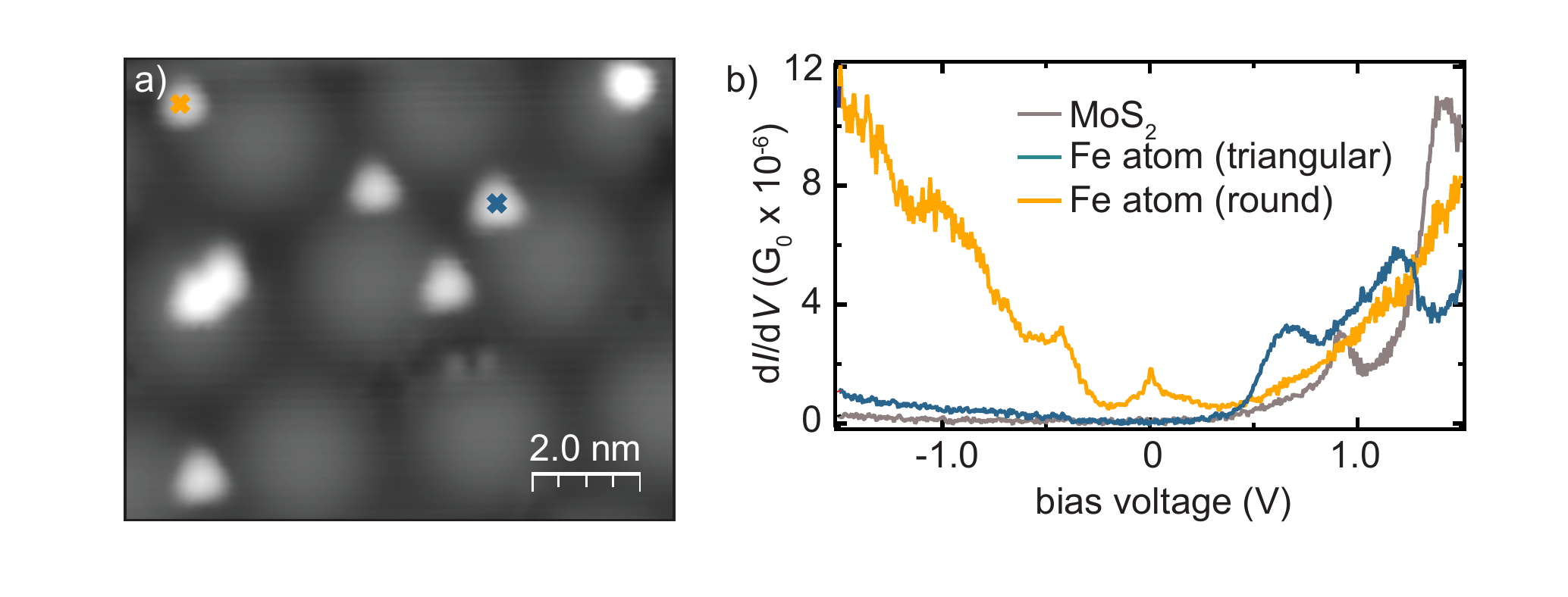}
\caption{a) Close-up view of Fe atoms adsorbed on a \mos\ island. The moir\'e structure is seen in the background. The Fe atoms appear either triangular or circular shaped. Image recorded at $V=\SI{50}{mV}$, $I=\SI{100}{\pA}$.  b) Differential conductance spectra recorded on a triangular (blue) and circular (orange) shaped Fe atom. The spectrum on \mos\ (gray) is shown for reference, revealing the well-known band gap with conduction band onset at 0.5\,V and valence band onset at -1.4\,V \cite{SMiwa2014,SGrubisic2015, SKrane2018a}. Spectra taken after opening the current feedback at a setpoint of $V=\SI{1.5}{V}$, $I=\SI{300}{\pA}$, lock-in modulation $V_\mathrm{rms}=\SI{5}{\mV}$.}
\label{triangular}
\end{figure}

As described in the main manuscript, the majority of Fe atoms are imaged as triangular shaped protrusions at low bias voltage. In contrast, a minority of protrusions is of circular shape (Fig.\,\ref{triangular}a). 
A clear distinction between the triangular and circular shaped Fe atoms can also be found in their differential conductance spectra in a larger energy range (Fig.\,\ref{triangular}b). The triangular atoms exhibit states above 0.5\,V, but no states within the occupied region. 
This is in contrast to the findings on the round-shaped Fe atoms, which show substantial conductance in a broad energy range below -0.2\,V. We also observe a narrow peak at zero bias, which may be associated to a Kondo resonance. 

\begin{figure}[t]
\includegraphics[width=\textwidth]{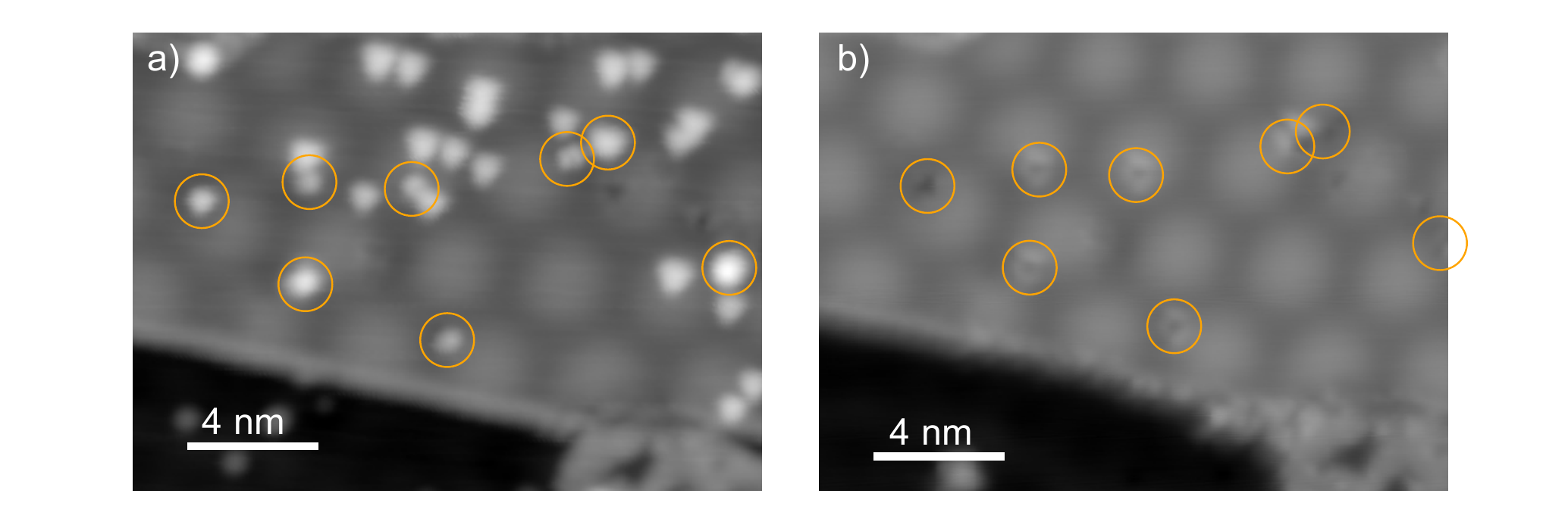}
\caption{a) STM topography image of an Fe-decorated \mos\ island. The majority of the Fe atoms appears triangular shaped, while some atoms take on different shapes (marked by orange circles). b) STM topography image of the same area as in (a) taken after removing all Fe atoms. This has been achieved by scanning at small bias voltage and large currents ($V=\SI{4.7}{mV}$, $I=\SI{25}{\nA}$). The locations of the atoms which had been imaged with non-triangular shape are marked again by orange circles. Topographies were recorded at a setpoint of $V=\SI{50}{mV}$, $I=\SI{100}{\pA}$. }
\label{fig:bare}
\end{figure}

To unravel the origin of the different types of Fe atoms, we remove all Fe atoms by scanning at small tip-sample distance ($I=\SI{25}{\nA}$, $V=\SI{50}{\mV}$). The subsequently recorded STM image reveals several local defects (Fig.~\ref{fig:bare}). Comparison to the previous STM image with the Fe atoms, we conclude that the round shaped Fe atoms had been located on defects. While we have shown an example of one Fe atom on a defect site in Fig.\,\ref{triangular}b, we remark that Fe atoms on different defects showed a large variety of \didv\ features which we disregard in this paper.

\section{Determination of the potential scattering $U$ at an Fe atom close to the moir\'e minimum}
\begin{figure*}[t]
\includegraphics[width=.95\textwidth]{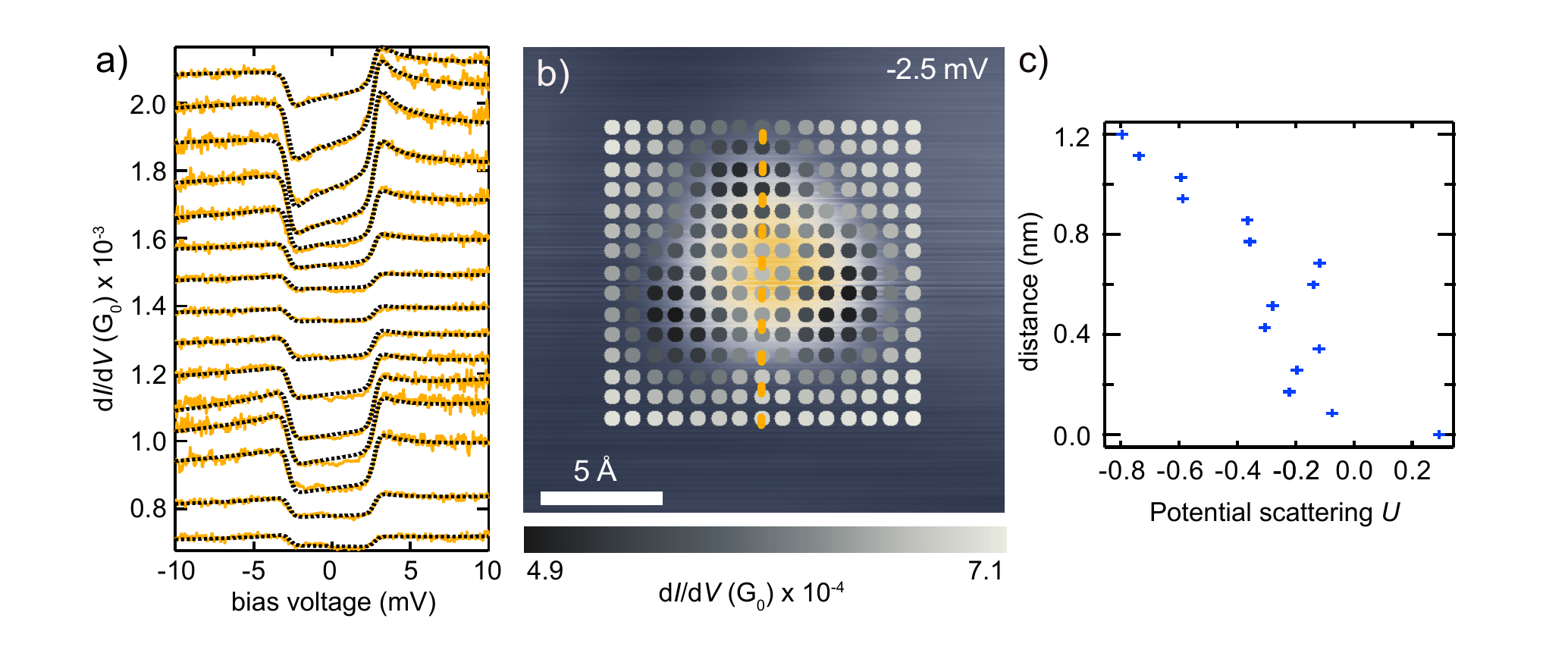}
\caption{a) Stacked plot of \didv\ spectra taken along the orange dashed line in (b), spectra are offset for clarity. Every second spectrum was also plotted in Fig. 3d of the main manuscript. b) Differential conductance signal extracted from a densely-spaced grid of spectra across an Fe atom on the moir\'e minimum at $V=\SI{-2.5}{mV}$ (same as in Fig. 3e in the main text). c) Extracted values for the potential scattering parameter $U$ obtained by fits according to Ref.\,\cite{STernes2015} with $J \rho =-0.132$ and $D=2.75$\,meV kept constant  (fits shown as black dashed lines in (a)). The spectra in (a) and (b) were recorded at a setpoint of $V=\SI{10}{mV}$, $I=\SI{1}{\nA}$, with a lock-in modulation of $V_\mathrm{rms}=\SI{50}{\uV}$.}
\label{fig:pot_scat}
\end{figure*}

In the main manuscript we discussed the differences between spectra recorded on and in the immediate vicinity of the Fe atoms. When the Fe atoms sit on or close to the minima of the moir\'e structure, the \didv\ lineshapes can be fitted by calculating the conductance including inelastic electron scattering up to second order in Born approximation and potential scattering \cite{STernes2015}. These fits allow for an evaluation of the variations in exchange and potential scattering strength. Here, we present a set of spectra taken along one of the high-symmetry axes of the Fe--S complex (spectra in Fig.\,\ref{fig:pot_scat}a along dashed line in \ref{fig:pot_scat}b). From the corresponding fits, we deduce that the exchange scattering $J\rho$ does not vary, while the potential scattering $U$ varies across the Fe--S complex (\ref{fig:pot_scat}c). We observe the largest values of $U$ at the triangle's vertex and a gradual decrease towards the center. 

\section{Local variations of lineshapes on Fe atoms close to the moir\'e maximum}

In the main manuscript we investigated the spatial variations of spectra in close vicinity of an Fe atom adsorbed on a minimum of the moir\'e structure. The reason for this choice was that the Fe atom was one of the most weakly coupled to the substrate such that the excitations could be described within second-order perturbation theory, and the lineshape could be fitted by the code developed by Ternes \cite{STernes2015}. In Fig.\,\ref{fig:maps} we provide complementary data on an Fe atom adsorbed at a rim towards a maximum of the moir\'e structure. As explained in the main manuscript, the \didv\ lineshape of the low-energy spectrum recorded on the center of the atom is highly asymmetric in bias voltage (blue spectrum in Fig.\,\ref{fig:maps}b). The lineshape at the vertex of the triangular shape is again different compared to the center (purple spectrum in Fig.\,\ref{fig:maps}b). The spectral variation along one of the (expected) threefold symmetry axes of the Fe--S complex is plotted in Fig.\,\ref{fig:maps}d.
The spatial variations are additionally mapped by plotting the \didv\ signal at $-0.9$\,mV and $1.6$\,mV from a densely-spaced grid of \didv\ spectra (Fig.\,\ref{fig:maps}e,f). Similar to the case of the Fe atoms on the moir\'e minima, the asymmetry is enhanced at the vertices of the Fe--S complex. However, the expected threefold symmetry of the Fe--S complex is broken by the superposition of the moir\'e modulation, which additionally contributes to the lineshape variations. 

To probe the correlation of the low-energy variations with the higher energy electronic structure, we plot the \didv\ spectra in a larger energy range in Fig.\,\ref{fig:maps}c. On the Fe center, the spectrum consists of a broad slope, while a spectrum on a vertex shows a faint peak at $\sim \SI{85}{mV}$. We extract its spatial intensity distribution from a densely-spaced grid of spectra (Fig.\,\ref{fig:maps}g). It reflects the same symmetry as the low-energy features in Fig.\,\ref{fig:maps}e,f, corroborating again the correlation between the orbital structure and lineshape of inelastic excitations. Although we cannot fit these spectra and therefore extract values of $U$, the asymmetry variations are in agreement with a variation of potential scattering as has been quantified from Fe atoms on the moir\'e minimum. We thus suggest that the variations in spectral features can be explained by the different contributions to the interfering tunneling paths.

\begin{figure*}[t]
\includegraphics[width=.95\textwidth]{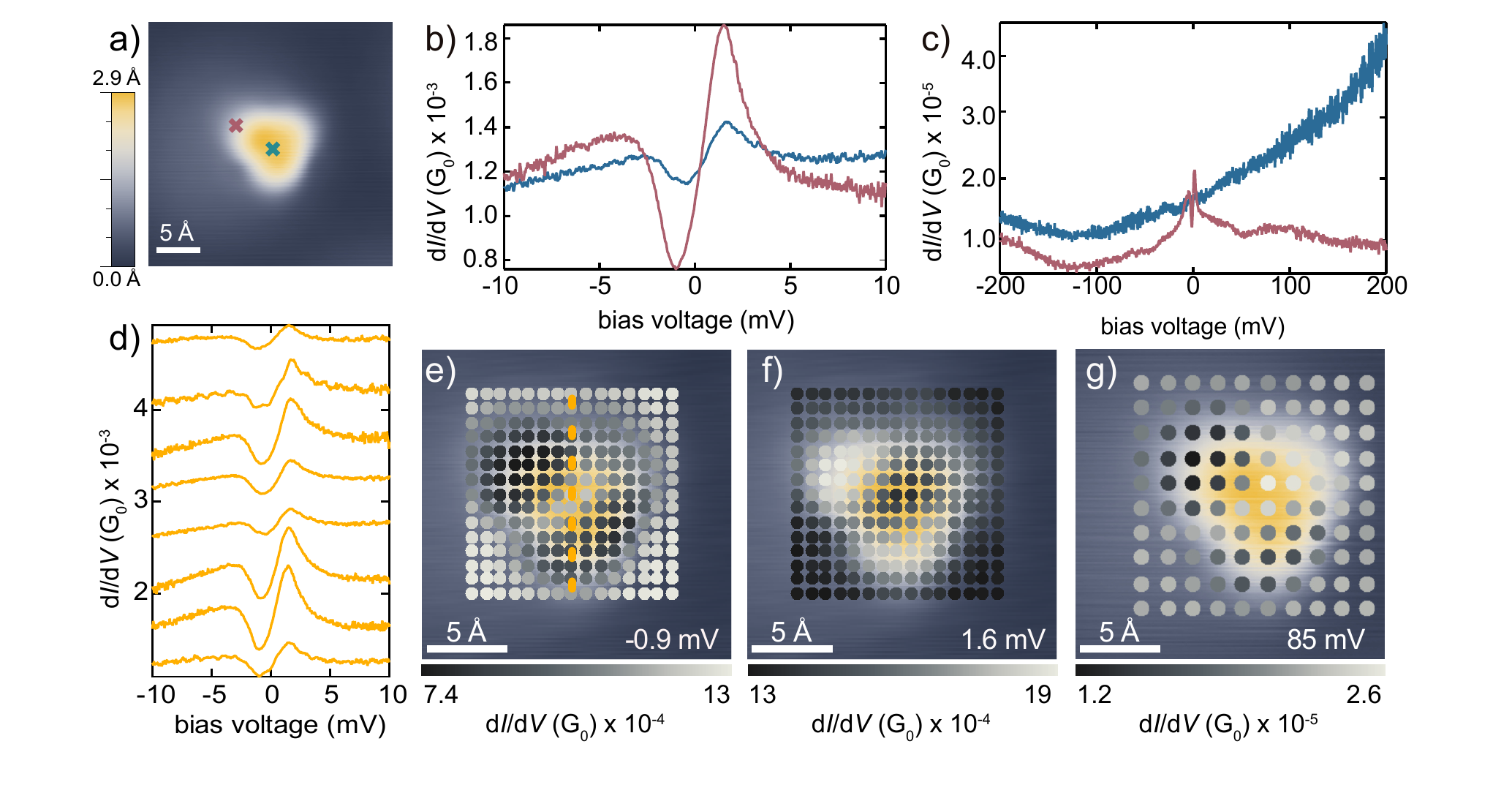}
\caption{ a) STM topography image of an Fe atom close to a maximum of the moir\'e structure ($V=\SI{10}{mV}$, $I=\SI{1}{\nA}$). b,c) Differential conductance spectra on the center (blue cross) and vertex (purple cross) of triangular shape of the Fe adatoms. d) Stacked plot of spectra (offset for clarity) along the orange dashed line indicated in e). e-g) STM topographies (blue-yellow, background) with superimposed differential conductance signal at the indicated bias voltage (black-white dots, scale given below panels) extracted from a densely spaced grid of spectra across the Fe atom at the indicated energies. The spectra in (b), (e) and (f) were recorded at a setpoint of $V=\SI{10}{mV}$, $I=\SI{1}{\nA}$, the spectra in (c) and (g) at $V=\SI{10}{mV}$, $I=\SI{20}{\pA}$, with an additional retraction of the tip by $z=\SI{20}{\pm}$. The lock-in modulation amplitude for (b), (e) and (f) was $V_\mathrm{rms}=\SI{50}{\uV}$, for (c) and (g) $V_\mathrm{rms}=\SI{1}{\mV}$.}
\label{fig:maps}
\end{figure*}

\bibliographystyle{apsrev4-2}

\end{document}